%% file: main.tex
\documentclass[11pt,a4paper]{article}
\usepackage[T1]{fontenc}
\usepackage[utf8]{inputenc}
\usepackage{lmodern}
\usepackage[margin=25mm]{geometry}
\usepackage{amsmath,amssymb,booktabs,graphicx,array,tabularx}
\usepackage[round,authoryear]{natbib}
\usepackage{placeins,microtype,xcolor}
\usepackage[colorlinks=true,linkcolor=blue!45!black,citecolor=blue!45!black,urlcolor=blue!45!black]{hyperref}

\newcommand{\unitE}{\mathrm{m^2\,Hz^{-1}}}
\input{numbers}

\input{physics_numbers}

\input{review_numbers}
\input{figures}
\title{Data-driven and physics-guided hindcasting of measured wave spectra}
\author{Sander Rikka$^{1,*}$\\[0.7em]
\small $^1$Department of Marine Systems, Tallinn University of Technology\\
\small $^*$Corresponding author: \href{mailto:sander.rikka@taltech.ee}{sander.rikka@taltech.ee}}
\date{} % remove date under author block
\begin{document}
\maketitle
\input{sections/abstract}
\input{sections/introduction}
\input{sections/methods}
\input{sections/results}
\input{sections/discussion}
\input{sections/conclusion}
\section*{Acknowledgements}
The wave-spectrum data were acquired through the Estonian Research Council Proof-of-Concept Grant EAG249. We thank the Finnish Meteorological Institute for providing the buoy spectra.
\section*{Data availability}
The buoy spectra were provided by the Finnish Meteorological Institute through project EAG249. The authors do not have permission to redistribute these spectra; access must be arranged with FMI. ERA5 forcing is from the Copernicus Climate Data Store product cited in \citet{era5dataset2018}. Bathymetry is from the Baltic Sea Bathymetry Database version 0.9.3, 500-m grid, downloaded on 3 December 2025  \citep{bsbd2013}. The coastline is from GSHHG version 2.3.7 at full resolution \citep{gshhg2017}. The code used for data preprocessing, model training and evaluation in this study is available from the corresponding author upon reasonable request.
\section*{Author contributions and competing interests}
Sander Rikka: conceptualization, modelling, manuscript writing and review. The author declares that they have no competing interests.
\FloatBarrier
\begingroup
\small
\bibliographystyle{plainnat}
\bibliography{references}
\endgroup
\end{document}

%% file: numbers.tex
\newcommand{\TrainCount}{73,793}
\newcommand{\ValCount}{23,862}
\newcommand{\TestCount}{23,179}
\newcommand{\FreqCount}{52}
\newcommand{\FreqStep}{0.01}
\newcommand{\FreqMin}{0.085}
\newcommand{\FreqMax}{0.595}

\newcommand{\AiiSpec}{0.40}

\newcommand{\AiiiHs}{18.78}
\newcommand{\AiiiAmp}{1.33}

\newcommand{\HeightPairCount}{278,148}
\newcommand{\AiiSpecGain}{2.63}
\newcommand{\AiiFpGain}{3.09}

\newcommand{\WindDecreaseCount}{7,614}
\newcommand{\WindIncreaseCount}{7,587}
\newcommand{\WindLowCount}{1,436}
\newcommand{\WindMidCount}{5,241}
\newcommand{\WindHighCount}{1,301}

%% file: physics_numbers.tex
\newcommand{\PhPtwoHs}{19.26}

\newcommand{\PhPthreeSpec}{0.39}
\newcommand{\PhPthreeHs}{19.66}

\newcommand{\PhPoneSpecChange}{8.44}
\newcommand{\PhPoneHsChange}{4.89}

\newcommand{\PhLaterPthreeSpec}{0.45}
\newcommand{\PhLaterPthreeHs}{24.87}

\newcommand{\PhEligible}{51,759}

\newcommand{\PhExtendedCount}{22,892}

\newcommand{\PhPtwoHsSkill}{51.55}
\newcommand{\PhPthreeSpecSkill}{36.35}

\newcommand{\PhPtwoHsBias}{-5.71}

\newcommand{\PhPtwoHsCorr}{0.97}

%% file: review_numbers.tex
\newcommand{\RvOriginalAiiFp}{$4.80\times10^{-2}$}
\newcommand{\RvLowAiiSpec}{0.08}
\newcommand{\RvHighAiiSpec}{0.81}
\newcommand{\RvLowAiiNspec}{0.81}
\newcommand{\RvHighAiiNspec}{0.49}
\newcommand{\RvIncPthreeSpecGain}{2.79}
\newcommand{\RvDecPthreeSpecGain}{1.59}
\newcommand{\RvHighPthreeSpecGain}{2.80}
\newcommand{\RvDecPthreeHsGain}{4.85}
\newcommand{\RvHighPthreeHsGain}{5.65}
\newcommand{\RvIncPoneSpecCost}{20.70}

%% file: figures.tex
\newcommand{\FigureHorizon}{%
\begin{figure}[htbp]
\centering\includegraphics[width=\textwidth]{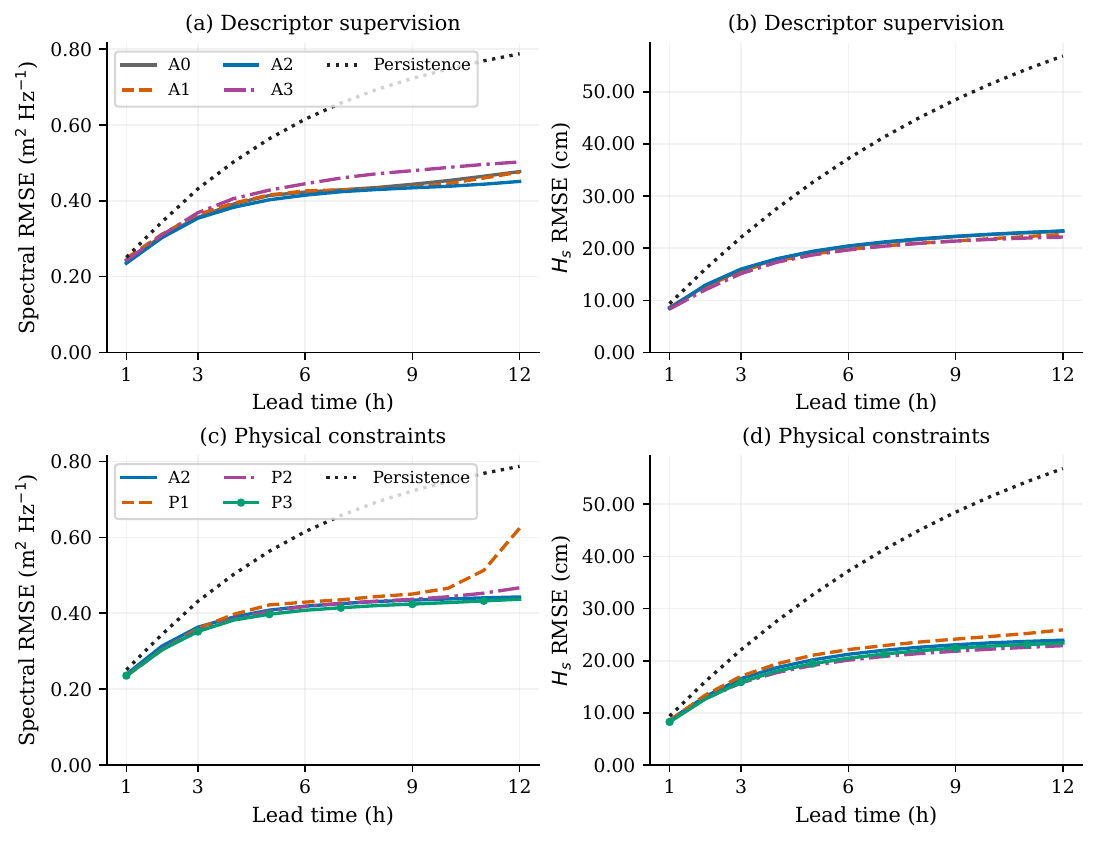}
\caption{Spectral and wave-height RMSE by lead time: descriptor objectives (a,b) and physical constraints (c,d). The lower panels use the matched A2 reference. All curves use the same 23,179 test origins; persistence repeats the initial spectrum.}\label{fig:horizon}
\end{figure}
}

\newcommand{\FigureWind}{%
\begin{figure}[htbp]
\centering\includegraphics[width=\textwidth]{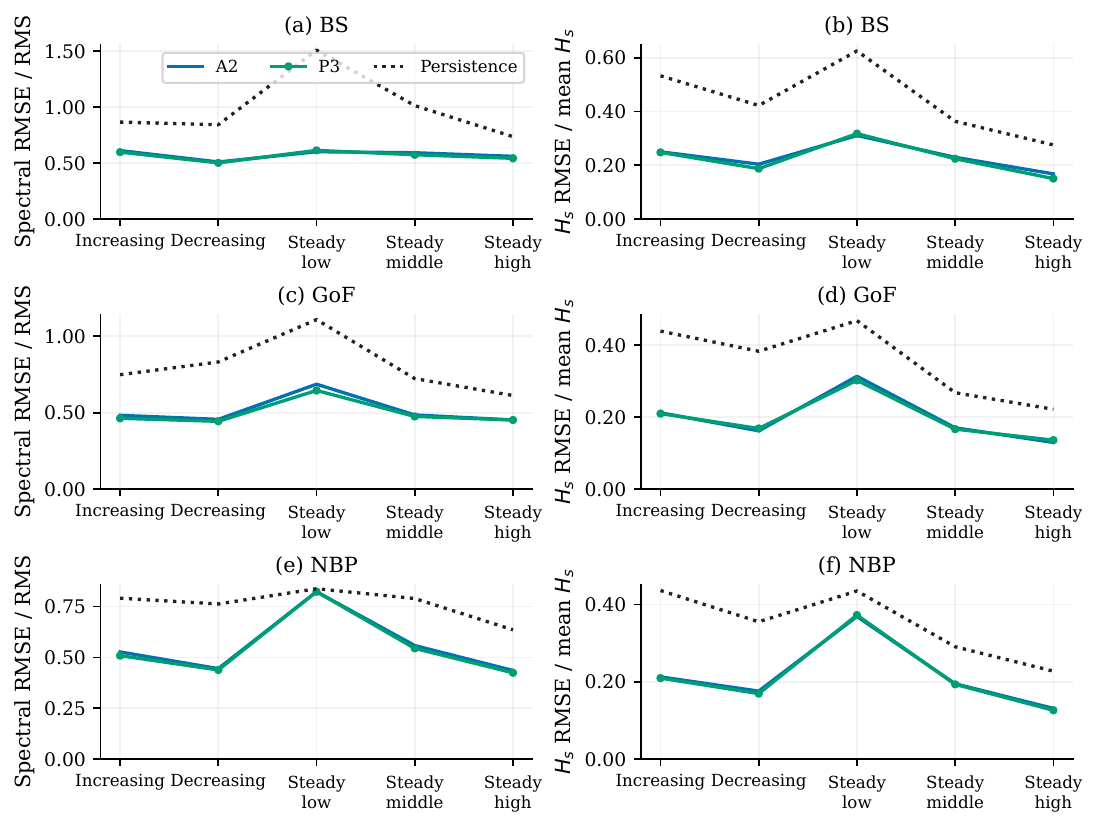}
\caption{Wind-conditioned normalized errors for the matched A2 reference, extended relaxation (P3) and persistence. Rows show BS, GoF and NBP. Each value pools leads 1--12 within its station and wind class. Steady denotes small net wind-speed change, subdivided by mean speed; denominators use observations within each subset.}\label{fig:wind}
\end{figure}
}

\newcommand{\FigureExamples}{%
\begin{figure}[htbp]
\centering\includegraphics[width=\textwidth]{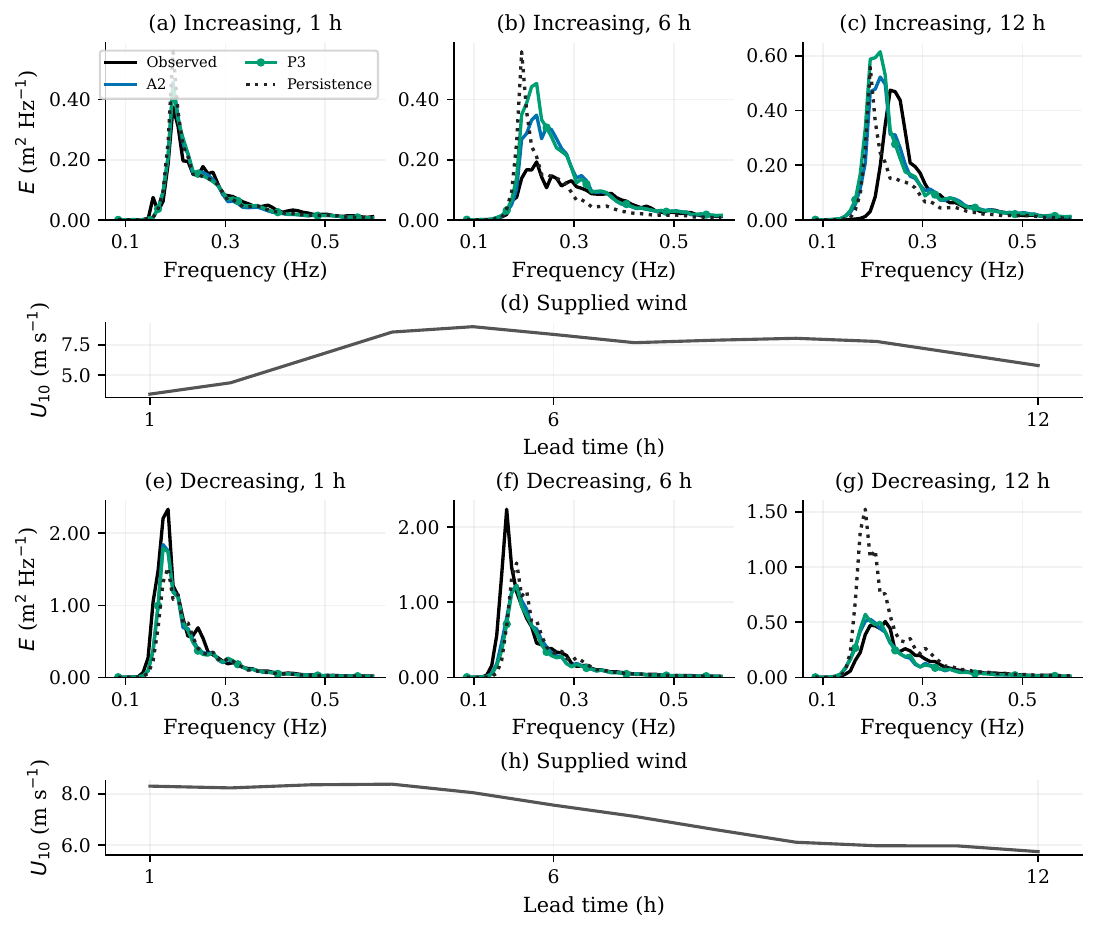}
\caption{NBP spectral hindcasts and supplied wind for two forcing-selected origins: 30 June 2022, 21:00 UTC (a--d), and 2 December 2022, 14:00 UTC (e--h). Spectral panels compare observations, the matched A2 reference, P3 and persistence at one, six and twelve hours. Cases are selected from wind-class medians without using prediction errors.}\label{fig:examples}
\end{figure}
}

\newcommand{\FigureExtension}{%
\begin{figure}[htbp]
\centering\includegraphics[width=\textwidth]{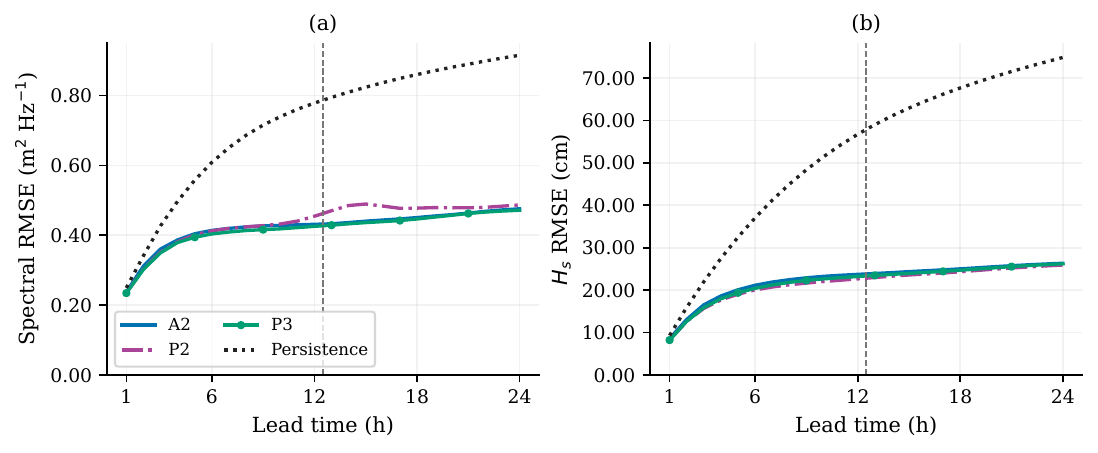}
\caption{Continuation to 24 hours for the A2 reference, relaxation (P2), extended relaxation (P3) and persistence on 22,892 common test origins. The dashed line marks the end of spectral supervision. P3 alone received a physical residual at hours 13--24 during training.}\label{fig:physicsextension}
\end{figure}
}

%% file: sections/abstract.tex
\begin{abstract}
Short-range hindcasts of measured wave spectra are evaluated using a recurrent model initialized by one hourly buoy spectrum and supplied with local wind, fetch and depth. Measurements from three Baltic Sea stations are combined for training in 2017--2020, validation in 2021 and testing in 2022. ERA5 reanalysis supplies winds throughout the twelve-hour prediction interval. The comparison examines supervision of spectral variance, peak frequency and peak amplitude, followed by empirical wind-sea bounds and a local relaxation constraint. Variance and peak-frequency supervision gives the lowest spectral error within the data-driven comparison, whereas additional peak-amplitude supervision gives the lowest wave-height error. In the matched physical comparison, relaxation reduces height RMSE by 5.01\% to \PhPtwoHs{}~cm. Extending this constraint through hours 13--24, without spectral targets at those hours, reduces primary spectral RMSE by 2.31\% to \PhPthreeSpec{}~$\unitE$. Upper-bound constraints increase errors. Wind-condition comparisons show spectral gains from extended relaxation in increasing, decreasing and steady strong winds, with little change in steady weak winds. The results support modest, objective-dependent benefits from local physical guidance. Wave-height scores concern the retained frequency band, and operational forecasting and transfer to independent stations remain to be evaluated.
\end{abstract}

%% file: sections/introduction.tex
\section{Introduction}
Reliable information on sea state is important for maritime safety, coastal management, offshore engineering and the planning of marine operations. Wave measurements describe the conditions encountered at a site, numerical models extend this information in space and time, and forecasts support decisions before adverse conditions develop. Both typical conditions and the intensity and duration of energetic events matter for these applications \citep{rutgersson2022,bjorkqvist2024events}. Improving the representation and short-term prediction of waves therefore remains relevant even in seas with a comparatively moderate mean wave climate.

The Baltic Sea is a semi-enclosed basin with an irregular coastline, distinct sub-basins and seasonal ice cover. Its wave climate is dominated by locally generated windsea, with limited penetration of ocean swell. The 41-year hindcast of \citet{bjorkqvist2018} gives open-sea ice-free mean significant wave heights broadly within $0.5$--$1.5$~m, with the highest values in the exposed Baltic Proper and lower values in sheltered sub-basins. Published buoy records include storm significant wave heights of $8.2$~m in the northern Baltic Proper, $8.1$~m in the Bothnian Sea and $5.2$~m in the Gulf of Finland \citep{bjorkqvist2018,bjorkqvist2020}. These are measured extremes reported in those studies; the 41-year hindcast gives a local maximum of $10.1$~m in the southern Baltic Proper. Mean peak periods are approximately $5.0$--$6.0$~s in the Baltic Proper and below $5.0$~s in the Gulf of Finland, with similar statistics in the Bothnian Sea; peak periods during severe Baltic Proper storms can reach approximately $10.0$--$14.0$~s \citep{bjorkqvist2018}.

Wave growth depends on both the available fetch length and the duration of the wind forcing. The longest possible Baltic fetch is about $700$~km, but coastlines, islands and ice strongly restrict the effective fetch for many wind directions \citep{bjorkqvist2018}. Changes in wind direction can therefore alter wave conditions rapidly. \citet{soomere2011} describe relatively short saturation and memory times for Baltic wave fields, while \citet{bjorkqvist2021swell} show that locally generated waves may persist after the wind weakens. Their swell partition depends on wave speed and wind--wave direction, and does not necessarily represent long-period oceanic swell. These conditions motivate examining the measured initial spectrum together with evolving local winds over a short prediction horizon.

Spectral wave models describe variance distributed over frequency and direction and its evolution through propagation, wind input, dissipation and wave interactions. SWAN, for example, solves a spectral action balance incorporating these processes \citep{booij1999}. Such models have been used extensively to characterize Baltic wave conditions \citep{bjorkqvist2018}. For operational applications, the Finnish Meteorological Institute produces the Copernicus Marine Baltic Sea Wave Analysis and Forecast product \citep{copernicusbal2020}. The cited product manual specifies WAM cycle~4.7 on an approximately one-nautical-mile grid, with hourly output and twice-daily production. MetCoOp-HARMONIE supplies winds for the first $66$~h, followed by ECMWF forcing; currents, sea level and ice information are supplied by the regional ocean forecast system \citep{copernicuspum2025}. \citet{aguiar2024} evaluated an earlier WAM configuration and reported Baltic operational wave-height errors, including the one-to-twelve-hour range. Significant wave height is an important integrated measure, but spectra with the same variance can differ in peak location, peak density and shape. Wave-height accuracy alone is therefore insufficient to assess spectrum prediction.

Machine learning provides a complementary approach to short-term wave prediction. LSTM-based methods have been applied to significant wave height using historical wave and meteorological measurements \citep{fan2020}. Observation-driven short-range prediction has also been combined with spatial surrogates and evaluated against operational numerical forecasts \citep{chen2023forecast}. Nevertheless, greater model complexity does not ensure higher accuracy: \citet{jiang2024} found that neural models did not outperform autoregression in their univariate wave-height comparison. Appropriate baselines and a clearly defined input protocol remain necessary.

Deep learning has also been used to predict the spectrum itself. \citet{song2023} modeled single-point directional spectra from ERA5 wind fields, and \citet{gao2025} combined local and remote wind inputs with numerical hindcast spectra as targets. \citet{lu2026} developed a global directional-spectrum prediction framework using latent compression and reconstruction. Observed spectra have also been used as learning targets: \citet{filoche2026} corrected operational directional forecasts to one-dimensional buoy spectra. These approaches differ in whether they emulate numerical spectra, post-process operational forecasts or predict the subsequent evolution of an observed spectrum. Similarly, reanalysis-conditioned experiments, including global wave-height modeling by \citet{wangjiang2024}, have different information constraints from forecasts driven only by data available at issue time.

Physical knowledge can guide learning through observed wave descriptors or through constraints on wave evolution. Physics-informed neural networks include governing-equation residuals in the loss \citep{raissi2019}; ocean-wave applications include phase-resolved reconstruction \citep{ehlers2025prf} and reduced energy balances for regional wave parameters \citep{wang2026regional}. Other methods supervise spectral energy, peak frequency and shape when reconstructing spectra from ship motions \citep{sun2025}. Such descriptor supervision constrains agreement with observations, whereas a dynamical residual constrains agreement with an assumed physical relation.

This study examines how these two forms of guidance affect short-range hindcasts of measured one-dimensional wave spectra. One buoy spectrum initializes a twelve-hour prediction, with local wind, directional fetch and depth as inputs. The comparison first adds supervision of variance, peak frequency and peak amplitude, and then tests empirical wind-sea bounds and a local relaxation equation. A further test extends physical guidance beyond the hours with spectral targets. ERA5 reanalysis supplies the winds over each prediction interval; the study therefore concerns retrospective hindcasting. Measurements from three Baltic Sea stations provide contrasting wave conditions within the same training and evaluation framework.

Section~\ref{sec:methods} describes the data, model and constraints. Section~\ref{sec:results} presents the spectral and wave-height errors, their station and wind dependence, and the continuation test. Section~\ref{sec:discussion} discusses the physical interpretation and application limits, followed by the conclusions in Section~\ref{sec:conclusion}.

%% file: sections/methods.tex
\section{Data and methods}\label{sec:methods}
\subsection{Measured spectra and hindcasting task}
The study combines records from the Northern Baltic Proper (NBP), Bothnian Sea (BS) and Gulf of Finland (GoF) stations (Table~\ref{tab:stations}). The instruments are Datawell Directional Waverider Mark 4 buoys operated by the Finnish Meteorological Institute (FMI). A single model is trained jointly on the three stations. At each hourly origin $t$, one measured variance-density spectrum $E_t(f)$ initializes predictions for $t+1,\ldots,t+12$~h.

The spectra were resampled to hourly resolution during data preparation and aligned with hourly winds. Incomplete records and windows crossing time gaps are excluded. Spectra are linearly interpolated onto \FreqCount{} frequencies between \FreqMin{} and \FreqMax{}~Hz, at nominal spacing \FreqStep{}~Hz. This interpolation does not explicitly preserve integrated variance. The same grid is used for observations, predictions and persistence.

The retained variance and corresponding significant wave height are
\begin{equation}
 m_0^{\mathcal B}=\sum_{k=1}^{F}E(f_k)\Delta f_k,\qquad
 H_s^{\mathcal B}=4\sqrt{m_0^{\mathcal B}} .
 \label{eq:height}
\end{equation}
Here $\mathcal B$ denotes the retained frequency grid. Endpoint widths equal the adjacent frequency difference; interior widths equal half the two-neighbour difference. Throughout the following text, $m_0$ and $H_s$ denote these band-limited quantities, $m_0^{\mathcal B}$ and $H_s^{\mathcal B}$. The resulting variance and wave height can therefore differ from those calculated using the full frequency range of the original spectra.

\input{tables/stations}

Training targets fall in 2017--2020, validation targets in 2021 and test targets in 2022. Every twelve-hour target window must lie within its assigned period. The resulting counts are \TrainCount{}, \ValCount{} and \TestCount{} origins, respectively. All three stations contribute to each period. Successive hourly origins overlap and do not represent independent events.

\subsection{Wind, fetch and depth}
The ten-metre wind components $u_{10},v_{10}$ are from ERA5 hourly single-level reanalysis \citep{hersbach2020,era5dataset2018}, extracted from the closest point on its $0.25^\circ$ grid without spatial interpolation. Winds at the origin and throughout the prediction interval are supplied to the model. These are retrospective reanalysis winds; operational wind-forecast errors are not represented.

Upwind fetch is measured along eight rays from each station to the first land intersection, using the full-resolution GSHHG 2.3.7 coastline \citep{gshhg2017}. The nearest $45^\circ$ sector is selected from meteorological wind direction, $\theta=\operatorname{atan2}(-u_{10},-v_{10})$ modulo $360^\circ$. Station depth is taken from the nearest coordinate in the 500-m Baltic Sea Bathymetry Database (BSBD) 0.9.3 grid, downloaded on 3 December 2025 \citep{bsbd2013}. Its supplied vertical reference is retained.

\subsection{Model and comparison design}\label{sec:descriptors}
Table~\ref{tab:objectives} summarizes the seven objectives. The data-driven comparisons add supervision of measured spectral descriptors. The physics-guided comparisons add empirical wind-sea bounds or a local relaxation ordinary differential equation (ODE). These physical terms represent reduced local assumptions; the spatial wave-action partial differential equation (PDE) and complete source balance are not imposed.

\input{tables/objectives}

All objectives use the same encoder--decoder long short-term memory (LSTM) architecture \citep{hochreiter1997}. The encoder has three layers and the decoder one layer, each with 128 hidden units. The encoder receives the initial spectrum, wind and fetch; the decoder receives the preceding spectrum, wind and fetch for the next hour, and station depth. Wind components, fetch in km and signed bathymetry in m are used without further standardization. Spectra are represented as $Y=\log_{10}\max(E,\epsilon)$, with $\epsilon=10^{-8}$. At each step, a predicted increment updates the spectrum:
\begin{equation}
 \widehat Y_h=\operatorname{clip}(\widehat Y_{h-1}+\Delta Y_h,-12,4),
 \qquad \widehat Y_0=Y_t .
\end{equation}
The model feeds its own prediction into the next step during both training and evaluation. Linear spectra are recovered as $10^{\widehat Y_h}$; descriptor losses use densities floored at $\epsilon$.

Let $\langle\cdot\rangle$ denote a mean over the applicable samples, hours and frequencies. For $N=12$ prediction hours, batch size $B$ and $F$ frequency bins, the baseline loss is
\begin{align}
 L_{\rm rec}&=\langle(\widehat Y-Y)^2\rangle,\\
 L_{\rm incr}&=\frac{1}{(N-1)BF}\sum_{b=1}^{B}\sum_{h=2}^{N}\sum_{k=1}^{F}
 a_h\left[(\widehat Y_{b,h,k}-\widehat Y_{b,h-1,k})
 -(Y_{b,h,k}-Y_{b,h-1,k})\right]^2,\\
 L_{\rm A0}&=L_{\rm rec}+0.3L_{\rm incr}.
\end{align}
In $L_{\rm incr}$, $a_h$ weights the squared error in the change from lead $h-1$ to $h$: $a_2=1.25$ for 1--2~h, $a_3=1.15$ for 2--3~h and $a_4=1.05$ for 3--4~h; $a_h=1$ for all later pairs.

Variance supervision uses
\begin{equation}
 L_{m_0}=\left\langle[\log_{10}\max(\widehat m_0,\epsilon)
 -\log_{10}\max(m_0,\epsilon)]^2\right\rangle .
\end{equation}
A differentiable peak frequency is extracted from both predicted and observed log spectra:
\begin{equation}
 \widetilde f_p(Y)=\sum_k f_k\frac{\exp(Y_k/T)}{\sum_j\exp(Y_j/T)},\quad T=0.15,
 \qquad L_{\rm soft}=\left\langle
 \left[\frac{\widetilde f_p(\widehat Y)-\widetilde f_p(Y)}{f_F-f_1}\right]^2\right\rangle .
\end{equation}
For peak amplitude $A_p=\max_k E(f_k)$, the loss is
$L_{A_p}=\sqrt{\langle(\widehat A_p-A_p)^2\rangle+10^{-12}}$.
The objectives are
\begin{equation}
 L_{\rm A1}=L_{\rm A0}+0.1L_{m_0},\quad
 L_{\rm A2}=L_{\rm A1}+0.2L_{\rm soft},\quad
 L_{\rm A3}=L_{\rm A2}+0.001L_{A_p}.
\end{equation}
These are observation-misfit terms. The amplitude term uses linear density, whereas variance and peak-frequency supervision use logarithmic or normalized quantities.

\subsection{Local physical constraints}\label{sec:physics}
\subsubsection{Equilibrium wind sea}
For an hourly transition, let $U_h^*$ and $X_h^*$ be the means of the endpoint wind speeds and upwind fetch distances. With positive water depth $d$, define $\chi=gX_h^*/(U_h^*)^2$ and $\delta=gd/(U_h^*)^2$. The Young--Verhagen growth relations \citep{young1996,karimpour2016} give equilibrium variance and peak frequency:
\begin{align}
 a&=\tanh(0.493\delta^{0.75}),\qquad b=\tanh(0.331\delta^{1.01}),\\
 \varepsilon_{\rm eq}&=0.00364[a\tanh(0.00313\chi^{0.57}/a)]^{1.74},\\
 \nu_{\rm eq}&=0.133[b\tanh(0.0005215\chi^{0.73}/b)]^{-0.37},\\
 m_{\rm eq}&=\varepsilon_{\rm eq}(U_h^*)^4/g^2,\qquad
 f_{\rm eq}=\nu_{\rm eq}g/U_h^* . \label{eq:yv}
\end{align}
The represented wind-sea bins are
$\mathcal W_h=\{f_k\in\mathcal B:f_k\geq0.9(0.13g/U_h^*)\}$.
To project equilibrium variance onto these bins, a JONSWAP shape is used \citep{hasselmann1973}:
\begin{align}
 J(q)&=q^{-5}\exp(-1.25q^{-4})\,3.3^{\exp[-(q-1)^2/(2\sigma^2)]},\\
 E_{\rm eq}(f)&=\frac{m_{\rm eq}}{f_{\rm eq}}
 \frac{J(f/f_{\rm eq})}{\int_{0.05}^{100}J(q)\,dq},\qquad
 m_{{\rm eq},h}^{\mathcal W}=\sum_{f_k\in\mathcal W_h}E_{\rm eq}(f_k)\Delta f_k .
 \label{eq:bandprojection}
\end{align}
Here $q=f/f_{\rm eq}$, and $\sigma=0.07$ below the peak and $0.09$ above it. The broad normalization range preserves the distinction between total equilibrium variance and its represented fraction.

\subsubsection{Applicability and calibration}
Physical penalties require $U_h^*\geq5~\mathrm{m\,s^{-1}}$, at least four wind-sea bins, endpoint wind-speed change no greater than $2.5~\mathrm{m\,s^{-1}}$ and wind turn no greater than $30^\circ$. At least half the initial observed variance must lie in the step's $\mathcal W_h$; this condition uses the origin observation at every step. Fetch, depth and forcing must also be valid.

Calibration uses only the first observed transition at each training origin; \PhEligible{} of \TrainCount{} transitions meet these conditions. Stationwise 95th percentiles of observed/equilibrium variance and soft peak-period ratios, multiplied by 1.001, set upper-bound tolerances. A station is excluded from the corresponding bound if its ratio exceeds 4 for variance or 1.5 for period. The largest accepted ratio, with a minimum of one, gives $c_E=2.52$ and $c_T=1.24$. These bounds apply at BS and NBP; GoF remains in supervised training but is excluded from P1's physical penalties.

The relaxation time is selected from $\{1,2,3,6,12,24,48\}$~h. The smallest value satisfying a training-residual 90th absolute percentile of at most 0.15 log units and RMSE at most 1.10 times the observed persistence-increment RMSE is $\tau=12$~h. This empirical choice controls how strongly an hourly transition approaches equilibrium.

\subsubsection{Bounds and relaxation}
Both endpoints are integrated over the same $\mathcal W_h$, with the observed spectrum used at $h=0$. P1 penalizes excess wind-sea variance and excess soft peak period:
\begin{align}
 r_{E,h}&=[\log_{10}\widehat m_h^{\mathcal W_h}
              -\log_{10}(c_E m_{{\rm eq},h}^{\mathcal W})]_+,\\
 r_{T,h}&=[\log_{10}(f_{\rm eq}/c_T)
              -\log_{10}\widetilde f_h^{\mathcal W_h}]_+,\\
 L_{\rm P1}&=L_{\rm A2}+0.1\langle r_E^2\rangle_{\mathcal G_E}
                         +0.1\langle r_T^2\rangle_{\mathcal G_T}.
\end{align}
Here $[x]_+=\max(x,0)$, the soft peak is restricted to $\mathcal W_h$, and $\mathcal G_E,\mathcal G_T$ denote the applicable steps and station masks.

P2 and P3 instead constrain the local relaxation equation $dm/dt=(m_{\rm eq}-m)/\tau$. Its exact solution over one hour, holding the equilibrium and frequency subset fixed, gives
\begin{align}
 m_h^*&=m_{{\rm eq},h}^{\mathcal W}
       +(\widehat m_{h-1}^{\mathcal W_h}-m_{{\rm eq},h}^{\mathcal W})e^{-\Delta t/\tau},\\
 r_h&=\log_{10}\widehat m_h^{\mathcal W_h}-\log_{10}m_h^*,\\
 L_{\rm P2}&=L_{\rm A2}+0.1\langle r_h^2\rangle_{\mathcal G,1:12},\\
 L_{\rm P3}&=L_{\rm A2}+0.1\left(
       \langle r_h^2\rangle_{\mathcal G,1:12}
       +\langle r_h^2\rangle_{\mathcal G,13:24}\right). \label{eq:relaxation}
\end{align}
Eligible steps at all three stations contribute to these losses. Each residual term is averaged separately over eligible steps, with zero contribution for an empty set. Variances are floored at $\epsilon$ before logarithms, and both predicted states receive gradients.

For P3, the model continues to hour 24 during training, while observed-spectrum targets stop at hour 12. Missing winds and continuation steps outside the training period are masked. All supervised training origins are retained, and validation remains a twelve-hour task.

\subsection{Training and evaluation}
AdamW uses batch size 256, initial learning rate $1.758\times10^{-3}$ and weight decay 0.05. The learning rate halves every 50 epochs. Training runs for at most 90 epochs, with early stopping after 12 epochs without improvement in validation reconstruction loss. Checkpoints minimize 2021 $L_{\rm rec}$. The architecture and optimization are fixed; the physical comparisons use a separately fitted A2 reference under the same protocol. This reference is reported distinctly from A2 in the descriptor comparison.

Persistence repeats the origin spectrum. Spectral RMSE pools squared errors across origins, leads and frequencies before taking the square root. Descriptor RMSE similarly pools origin--lead pairs. Evaluation uses the hard peak $f_p=f_{\arg\max_k E(f_k)}$, peak amplitude and $H_s$. Wave-height bias is prediction minus observation; $r$ is Pearson correlation and $\mathrm{SI}=\mathrm{RMSE}/\overline{H_s}$. The primary aggregate covers leads 1--12, with \HeightPairCount{} overlapping origin--lead pairs.

Normalized spectral error divides RMSE by the root mean square (RMS) observed spectral density in the same subset. Normalized height error divides RMSE by mean observed $H_s$. Persistence skill is the percentage RMSE reduction,
\begin{equation}
 S_q=100\left(1-\frac{\mathrm{RMSE}_{q,\mathrm{model}}}
                         {\mathrm{RMSE}_{q,\mathrm{persistence}}}\right)\%.
\end{equation}
For continuation, A2, P2 and P3 are evaluated without further fitting on \PhExtendedCount{} origins having complete 24-hour observations. Hours 1--12 and 13--24 are scored separately on this common subset.

\subsection{Wind-condition analysis}
Each twelve-hour test interval is classified from ERA5 speed $U_h=\sqrt{u_h^2+v_h^2}$, its mean $\overline U$, and net gradient $G=(U_{12}-U_1)/(11~\mathrm{h})$. Wind is increasing when $G>0.1~\mathrm{m\,s^{-1}\,h^{-1}}$ and decreasing when $G<-0.1$. Remaining intervals are divided into low, middle and high speed using the training 20th and 80th percentiles, 4.4 and 9.7~$\mathrm{m\,s^{-1}}$. Classification uses the unrounded thresholds; equality enters the low or high group. Small net trend does not imply constant wind, and the increasing/decreasing classes need not be monotonic.

Errors and normalization denominators are calculated within each station and wind class. Two NBP examples are selected by proximity to the respective class medians of initial $H_s$, mean wind speed and net gradient, scaling each variable by its interquartile range. Selection does not use prediction errors.

%% file: tables/stations.tex
\begin{table}[htbp]
\centering\small
\caption{Buoy locations, water depths and aligned prediction-origin counts.}\label{tab:stations}
\begin{tabular}{lrrrrrr}
\toprule
Station & Latitude ($^\circ$N) & Longitude ($^\circ$E) & Depth (m) & Train & Validation & Test\\
\midrule
BS & 61.8000 & 20.2333 & 112.2 & 23,844 & 8,515 & 8,633\\
GoF & 59.9650 & 25.2350 & 49.7 & 22,264 & 6,794 & 6,851\\
NBP & 59.2500 & 21.0000 & 107.0 & 27,685 & 8,553 & 7,695\\
\bottomrule
\end{tabular}
\end{table}

%% file: tables/objectives.tex
\begin{table}[htbp]
\centering\small
\caption{Comparison objectives. All variants share the same inputs and network.}\label{tab:objectives}
\begin{tabularx}{\textwidth}{llX}
\toprule
ID & Descriptive name & Objective\\
\midrule
A0 & Spectrum baseline & Log-spectrum reconstruction and temporal increments\\
A1 & Variance supervision & A0 plus observed spectral variance\\
A2 & Peak-frequency supervision & A1 plus observed soft peak frequency\\
A3 & Peak-amplitude supervision & A2 plus observed peak spectral density\\
\midrule
P1 & Wind-sea bounds & A2 plus upper bounds on wind-sea variance and peak period\\
P2 & Relaxation & A2 plus a local relaxation residual at hours 1--12\\
P3 & Extended relaxation & P2 plus the residual at hours 13--24, without spectral targets\\
\bottomrule
\end{tabularx}
\end{table}

%% file: sections/results.tex
\section{Results}\label{sec:results}
\subsection{Spectral and wave-height accuracy}
All learned models improve pooled spectral, wave-height, peak-frequency and peak-amplitude RMSE relative to persistence (Table~\ref{tab:test}). Within the descriptor comparison, variance and peak-frequency supervision (A2) gives the lowest spectral error, \AiiSpec{}~$\unitE$, and peak-frequency error, \RvOriginalAiiFp{}~Hz. These errors are \AiiSpecGain{}\% and \AiiFpGain{}\% lower than the baseline. Adding peak-amplitude supervision (A3) instead gives the lowest height and amplitude errors, \AiiiHs{}~cm and \AiiiAmp{}~$\unitE$, respectively. Its spectral error is higher. Spectral shape, integrated variance and peak amplitude therefore favour different objectives.

\input{tables/test}

Physical constraints also produce mixed effects. Relative to the matched A2 reference, wind-sea bounds (P1) increase spectral RMSE by \PhPoneSpecChange{}\% and height RMSE by \PhPoneHsChange{}\%. Relaxation over twelve hours (P2) reduces height RMSE by 5.01\% to \PhPtwoHs{}~cm, with a small increase in spectral error. Extended relaxation (P3) gives the lowest spectral RMSE in this comparison, \PhPthreeSpec{}~$\unitE$, reducing spectral and height errors by 2.31\% and 3.04\%. The A2 reference retains the lowest peak-frequency error.

P3 spectral skill relative to persistence is \PhPthreeSpecSkill{}\%, while P2 height skill is \PhPtwoHsSkill{}\%. All learned models underestimate $H_s$ on average. P2, for example, has a bias of \PhPtwoHsBias{}~cm and correlation $r=\PhPtwoHsCorr{}$. Errors generally increase with lead time, but more slowly than persistence (Fig.~\ref{fig:horizon}).

\FigureHorizon

\subsection{Station differences}
Table~\ref{tab:stationerrors} compares A2 and P3 by station. GoF has the smallest absolute spectral error, but its normalized error is close to that at NBP. The absolute contrast partly reflects lower observed spectral density at GoF. P3 reduces both spectral and height errors at BS and NBP; at GoF, spectral error decreases slightly while height error increases. Both models retain positive persistence skill at all three stations.

\input{tables/station_errors}

\subsection{Dependence on wind conditions}
The test contains \WindIncreaseCount{} increasing-wind, \WindDecreaseCount{} decreasing-wind and \WindLowCount{}, \WindMidCount{} and \WindHighCount{} low-, middle- and high-speed small-trend origins. Absolute spectral errors are lowest in the low-speed class. For the A2 reference they range from \RvLowAiiSpec{}~$\unitE$ in that class to \RvHighAiiSpec{}~$\unitE$ in the high-speed class, whereas normalized errors are \RvLowAiiNspec{} and \RvHighAiiNspec{}. Small absolute errors in weak seas thus coexist with substantial relative errors (Fig.~\ref{fig:wind}).

\FigureWind

P3 reduces pooled spectral error relative to A2 by \RvIncPthreeSpecGain{}\% in increasing winds, \RvDecPthreeSpecGain{}\% in decreasing winds and \RvHighPthreeSpecGain{}\% in the high-speed small-trend class. Its largest height reductions among these classes occur in decreasing and high-speed winds, at \RvDecPthreeHsGain{}\% and \RvHighPthreeHsGain{}\%. In low-speed small-trend conditions the spectral change is below 0.1\% and height error increases slightly. P1 is less favourable: spectral error increases by \RvIncPoneSpecCost{}\% during increasing winds. The physical constraints therefore do not improve every wind class.

Figure~\ref{fig:examples} shows the A2 reference and P3 at one, six and twelve hours for two NBP origins selected from the increasing- and decreasing-wind classes. The observed spectra, persistence and supplied winds show how peak placement and density evolve within each interval. These cases illustrate the predictions without establishing general event-level performance.

\FigureExamples

\subsection{Continuation beyond twelve hours}
On the common 24-hour subset, P3 gives spectral RMSE \PhLaterPthreeSpec{}~$\unitE$ and height RMSE \PhLaterPthreeHs{}~cm at hours 13--24 (Fig.~\ref{fig:physicsextension}). Compared with A2, these are reductions of 0.76\% and 1.18\%. Compared with P2, P3 reduces spectral error by 6.30\%, while height error increases by 1.46\%. All three models outperform persistence. Extending the physical loss beyond the spectral targets therefore provides a modest spectral benefit, with a height tradeoff.

\FigureExtension
\FloatBarrier

%% file: tables/test.tex
\begin{table}[htbp]
\centering\small
\setlength{\tabcolsep}{4pt}
\caption{Pooled errors over leads 1--12~h. The two A2 rows are separate fits of the same supervised objective, one for each comparison. The WAM row provides regional context from \citet{aguiar2024}, rather than a matched benchmark.}\label{tab:test}\label{tab:heightdomain}\label{tab:physicsprimary}
\begin{tabular}{lrrrrrrr}
\toprule
 & \multicolumn{3}{c}{Spectral RMSE} & \multicolumn{4}{c}{Wave-height metrics}\\
\cmidrule(lr){2-4}\cmidrule(lr){5-8}
Model & $E$ ($\unitE$) & $f_p$ (Hz) & $A_p$ ($\unitE$) & RMSE (cm) & Bias (cm) & $r$ & SI\\
\midrule
\multicolumn{8}{l}{\emph{Descriptor comparison}}\\
A0 & 0.41 & $4.95\times10^{-2}$ & 1.60 & 19.54 & -6.26 & 0.97 & 0.20\\
A1 & 0.41 & $4.90\times10^{-2}$ & 1.48 & 19.00 & -3.98 & 0.97 & 0.19\\
A2 & 0.40 & $4.80\times10^{-2}$ & 1.56 & 19.62 & -5.08 & 0.97 & 0.20\\
A3 & 0.43 & $4.88\times10^{-2}$ & 1.33 & 18.78 & -4.32 & 0.97 & 0.19\\
\midrule
\multicolumn{8}{l}{\emph{Physical comparison}}\\
A2 (reference) & 0.40 & $4.80\times10^{-2}$ & 1.61 & 20.28 & -6.64 & 0.97 & 0.21\\
P1 & 0.43 & $4.83\times10^{-2}$ & 1.80 & 21.27 & -8.02 & 0.96 & 0.22\\
P2 & 0.40 & $4.82\times10^{-2}$ & 1.57 & 19.26 & -5.71 & 0.97 & 0.20\\
P3 & 0.39 & $4.80\times10^{-2}$ & 1.46 & 19.66 & -6.08 & 0.97 & 0.20\\
\midrule
Persistence & 0.61 & $6.61\times10^{-2}$ & 2.37 & 39.76 & -0.11 & 0.85 & 0.40\\
\midrule
Aguiar WAM & -- & -- & -- & 19.00 & 2.00 & 0.97 & 0.19\\
\bottomrule
\end{tabular}
\end{table}

%% file: tables/station_errors.tex
\begin{table}[htbp]
\centering\small
\caption{Station errors for the matched A2 reference and extended relaxation (P3). $S$ is RMSE reduction relative to persistence.}\label{tab:stationerrors}
\begin{tabular}{llrrrrrr}
\toprule
 & & \multicolumn{3}{c}{Spectral density} & \multicolumn{3}{c}{Wave height}\\
\cmidrule(lr){3-5}\cmidrule(lr){6-8}
Station & Model & RMSE ($\unitE$) & RMSE/RMS & $S$ (\%) & RMSE (cm) & SI & $S$ (\%)\\
\midrule
BS & A2 & 0.48 & 0.56 & 33.06 & 22.54 & 0.23 & 49.69\\
BS & P3 & 0.47 & 0.54 & 34.55 & 21.47 & 0.22 & 52.10\\
GoF & A2 & 0.14 & 0.47 & 37.77 & 13.87 & 0.18 & 50.11\\
GoF & P3 & 0.14 & 0.45 & 39.29 & 13.96 & 0.18 & 49.78\\
NBP & A2 & 0.45 & 0.47 & 36.65 & 22.30 & 0.19 & 47.74\\
NBP & P3 & 0.44 & 0.46 & 38.16 & 21.77 & 0.19 & 48.98\\
\bottomrule
\end{tabular}
\end{table}

%% file: sections/discussion.tex
\section{Discussion}\label{sec:discussion}
\subsection{What physical guidance contributes}
The main result is a tradeoff between the quantities constrained during training. Supervision of variance and peak frequency favours spectral accuracy, whereas peak-amplitude supervision favours height and amplitude. The physical comparisons give a similar distinction: local relaxation improves height, and its extension beyond labelled hours improves spectral accuracy. Empirical upper bounds perform worse overall. Additional physical guidance is therefore useful only to the extent that its assumptions and weighting suit the observed wave evolution.

This study connects spectrum-prediction research \citep{song2023,gao2025,lu2026} with learning from in situ spectra \citep{filoche2026}. Its contribution is the controlled comparison of spectral descriptors and local physical constraints for the subsequent evolution of a measured buoy spectrum. Related descriptor supervision has been used in inverse spectrum reconstruction \citep{sun2025}; here, it is tested alongside empirical growth relations and a dynamical relaxation penalty in a time-evolving application.

The equilibrium relation links wind, fetch and depth, but several assumptions separate it from the measured state. Local depth approximates the bathymetry along the fetch. An instantaneous fetch and a small endpoint wind change do not establish that forcing has persisted long enough for equilibrium. The frequency cutoff and fixed JONSWAP projection cannot resolve crossing directional systems. These limitations matter in the Baltic Sea, where wind direction changes the effective fetch and waves may persist after the generating wind changes \citep{soomere2011,bjorkqvist2021swell}. A swell label based on wave age and wind--wave direction should not be equated with long-period oceanic swell.

Calibration allows departures from the estimated equilibrium; it does not independently validate that equilibrium. The bound tolerances and station exclusions are derived from training observations, and the 12-hour relaxation time is selected by residual criteria rather than measured as a physical adjustment time. Indeed, the selected relaxation closure alone does not outperform hourly persistence for eligible training transitions. Its usefulness is as a constraint on the learned spectra. The effects of the calibration rules, station masks and loss weights have not been isolated, so the present results support this particular formulation rather than universal parameter choices.

\subsection{Wind dependence and regional context}
The wind-class results add physical context to the pooled errors. P3 improves spectral accuracy in increasing, decreasing and steady strong winds, while its effect in steady weak winds is very small. This pattern is consistent with stronger physical guidance during sufficiently windy conditions, but the diagnostic classes are not the same as the stepwise applicability masks. Initial spectra, seasons and station mixtures also differ between classes. The comparison is descriptive and cannot identify the physical penalty as the cause of a regime-specific improvement.

Station normalization is equally important. Lower absolute errors at GoF partly reflect its lower spectral energy scale. Differences in wave climate, sheltering and fetch may also affect prediction difficulty, but their individual roles are not isolated. Supplying depth and fetch can encode station differences even without an explicit station identifier.

For regional context, \citet{aguiar2024} report operational WAM height bias of $2.00$~cm, RMSE $19.00$~cm, $r=0.97$ and SI $0.19$ at leads 1--12~h. The present height RMSEs are \PhPtwoHs{}~cm for P2, \PhPthreeHs{}~cm for P3 and \AiiiHs{}~cm for A3, although WAM has a smaller bias magnitude. Differences in forcing, frequency support and evaluation datasets prevent a matched performance ranking. Nevertheless, the similar reported height errors are encouraging given the simpler representation: the local approach uses one measured one-dimensional spectrum and local forcing, while operational WAM resolves directional spectra, propagation and source processes over a spatial grid.

\subsection{Scope and further application}
The retrospective wind information is a central limitation for operational use. A forecasting evaluation would require winds available at issue time and the actual availability time of the initial measured spectrum. Persistence establishes improvement over an unchanged sea state; a matched numerical-wave-model comparison and a stronger statistical baseline would be needed to establish practical advantages \citep{jiang2024}.

Across 121,800 distinct station-hours eligible for the study windows, trapezoidal integration of native 0.025--1.0~Hz spectra and their retained portion gave a median variance retention of 96.95\% and a median $H_s$ reduction of 1.27~cm (95th percentile: 2.65~cm). On the same retained interval, regridding changed variance by a median absolute 0.52\%, while the full endpoint weighting in Eq.~\ref{eq:height} increased variance by a median 0.14\% relative to trapezoidal integration. The 52-frequency grid thus retains most measured variance in typical conditions, although larger fractional losses occur in weak seas and some energetic NBP cases.

Directional information and spatial propagation are also absent. The principal comparisons use one fitted realization per objective, and overlapping origins repeatedly score related sea states; small differences therefore should not be interpreted as statistically established improvements. Although training and checkpoint selection use the stated earlier years, 2022 informed earlier development, so the physical comparisons are exploratory.

The same prediction structure could potentially accept spectra estimated by another instrument, satellite retrieval or numerical model. Differences in frequency support, averaging and observation error would first need to be accommodated and tested. Independent locations and operational forcing would then determine where a local measured-spectrum approach can complement existing wave-modelling systems.

%% file: sections/conclusion.tex
\section{Conclusion}\label{sec:conclusion}
Measured wave spectra provide a useful initial state for twelve-hour hindcasting with local wind and station information. All tested objectives improve spectral and wave-height errors relative to persistence, but the most accurate objective depends on the quantity of interest. Variance and peak-frequency supervision favours spectral accuracy; peak-amplitude supervision favours wave height and peak density.

Physical guidance produces modest and mixed changes. Local relaxation improves wave-height accuracy, while extending the relaxation constraint beyond the labelled interval improves spectral accuracy. Empirical wind-sea bounds worsen the pooled results. Wind-condition comparisons show that these effects vary with the forcing and wave-energy scale.

The method represents a local, reanalysis-conditioned application of data-driven and physics-guided learning. Its calibration assumptions, band-limited spectra and evaluation at the training locations constrain the interpretation. Operational winds, independent locations and matched numerical-model comparisons are needed before extending the conclusions to practical forecasting.